\documentclass[11pt,twoside]{article}
\usepackage{./asp2014}

\aspSuppressVolSlug
\resetcounters

\begin{document}

\title{Verification of commercial motor performance for WEAVE at the William Herschel Telescope}
\author{James~Gilbert,$^1$ Gavin~Dalton,$^{1,2}$ and Ian~Lewis$^1$
\affil{$^1$Department of Physics, University of Oxford, Oxford, UK}
\affil{$^2$RALSpace, STFC Rutherford Appleton Laboratory, Didcot, UK}}

\paperauthor{James Gilbert}{james.gilbert@physics.ox.ac.uk}{0000-0001-5065-2101}{University of Oxford}{Department of Physics}{Oxford}{Oxfordshire}{OX1 3RH}{UK}
\paperauthor{Gavin Dalton}{gavin.dalton@physics.ox.ac.uk}{}{University of Oxford}{Department of Physics}{Oxford}{Oxfordshire}{OX1 3RH}{UK}
\paperauthor{Ian Lewis}{ian.lewis@physics.ox.ac.uk}{}{University of Oxford}{Department of Physics}{Oxford}{Oxfordshire}{OX1 3RH}{UK}

\begin{abstract}
WEAVE is a 1000-fiber multi-object spectroscopic facility for the 4.2~m William Herschel Telescope.  It will feature a double-headed pick-and-place fiber positioning robot comprising commercially available robotic axes.  This paper presents results on the performance of these axes, obtained by testing a prototype system in the laboratory.  Positioning accuracy is found to be better than the manufacturer's published values for the tested cases, indicating that the requirement for a maximum positioning error of 8.0~microns is achievable.  Field reconfiguration times well within the planned 60~minute observation window are shown to be likely when individual axis movements are combined in an efficient way.
\end{abstract}

\section{Introduction}
The WHT Enhanced Area Velocity Explorer (WEAVE) is a next generation wide field fiber-fed multi-object spectroscopic facility for the 4.2~m William Herschel Telescope (WHT).  Details on its design and motivation are discussed in \citet{jgilbert_weave2014}.

The WEAVE fiber positioner system will enable the deployment of 1000 individual optical fibers or 20 integral field units across a flat focal plane 410~mm in diameter (2\deg~on sky).  The design employs a pick-and-place robotic gantry system with exchangeable field plates, a format chosen in light of its successful implementation in the 400-fiber 2dF instrument at the Anglo-Australian Telescope \citep{jgilbert_2df2002}.

WEAVE advances the 2dF concept with the addition of a second robot in order to reduce configuration times and thus allow full field reconfigurations within a nominal 60 minute observation window.  WEAVE will use commercial off-the-shelf (COTS) robotic axes as a means to minimize costs and development time.  The performance of these devices with regard to the instrument requirements is the focus of this paper.

\subsection{Fiber positioner requirements}
The performance requirements for the WEAVE positioner are shown in Table~\ref{table_requirements}.  The requirement on accuracy flows down from a top-level tolerance of 0.2~arcmin for fiber-target misalignment, split equally between the optical and positioner subsystems.  Given the nominal observation window of 60~minutes, an average move time of 3.4 seconds is calculated with the assumptions that i) every fiber must be parked at the perimeter of the field before repositioning, and ii) the use of two robots is 90\% efficient.

\begin{table}[!ht]
\caption{Top level fiber positioner requirements}
\smallskip
\begin{center}
{\small
\begin{tabular}{llll}
\tableline
\noalign{\smallskip}
    Specification                                   & Requirement               \\
\noalign{\smallskip}
\tableline
\noalign{\smallskip}
    Total allowable fiber-target misalignment (RMS) & $\leq$0.2" (11.4~\micron) \\
    \hskip 1eM Contribution from prime focus optics & $\leq$8.0~\micron         \\
    \hskip 1eM Contribution from positioner         & $\leq$8.0~\micron         \\
    Field reconfiguration time                      & $\leq$60 min              \\
    \hskip 1em Implied average time per move        & $\leq$3.4 s               \\
\noalign{\smallskip}
\tableline\
\end{tabular}
\label{table_requirements}
}
\end{center}
\end{table}

\subsection{Positional error sources}
The achieved positioning accuracy will be the sum of various distinct error sources.  Table~\ref{table_errors_estimate} lists the most significant ones, combined as the root of sum of squares (RSS) for simplicity.  This gives a total `bottom-up' error estimate of 9.0~microns when taking axis performance as the manufacturer's specification.  However, this reduces to 6.0~microns with the actual measured values presented in the following sections.

\begin{table}[!ht]
\caption{Bottom-up analysis of RMS positioning errors for the most significant sources, showing best estimates before and after testing}
\smallskip
\begin{center}
{\small
\begin{tabular}{lll}
\tableline
\noalign{\smallskip}
    Error source                    & Initial estimate~(\micron)    & Revised estimate~(\micron)    \\
\noalign{\smallskip}
\tableline
\noalign{\smallskip}
    Gantry non-orthogonality calibration    & 4.0                   & 4.0                           \\
    Metrology camera calibration            & 1.2                   & 1.2                           \\
    Fiber measurement error (0.2~pixels)    & 0.75                  & 0.75                          \\
    Robot axis accuracy ($x$+$y$)           & 7.1                   & 3.6                           \\
    Gripper jaws release repeatability      & 3.0                   & 0.91                          \\
    Gantry flexure variability              & 2.0                   & 2.0                           \\
\noalign{\smallskip}
\tableline
\noalign{\smallskip}
    RSS total                               & 9.0                   & 6.0                           \\
\noalign{\smallskip}
\tableline\
\end{tabular}
\label{table_errors_estimate}
}
\end{center}
\end{table}

\section{Prototype positioner robot}
A prototype positioner has been assembled in order to verify the performance of the COTS motors chosen for WEAVE.  It is a simplified version of the full design, matching the format of the $y$-, $z$-, $\theta$-axes and gripper assembly.  This allows pick-and-place positioning of magnetic fiber terminals, called `buttons', along a steel strip spanning the $y$-axis.  Metrology is provided by a camera within the gripper assembly.

\subsection{Performance tests}
The following sections present the results of positioning accuracy and speed tests for the prototype robot described above.  This initial evaluation of the chosen COTS motors indicates accuracies better than published values, and an estimated positioning time well within the maximum 60~minutes.

\altsubsubsection{Linear axis repeatability}
Axis accuracy was measured with a simple test of positional repeatability.  The robot was instructed to position 500~mm away from its origin and then return.  This routine was repeated 500 times and the apparent location of a static reference fiber extracted for each cycle.  The scatter in this measurement provides an upper limit for positioning accuracy of the axis.

The results (Figure~\ref{plot_axis_rep}) show a scatter of only 2.55~microns along the active axis, which is substantially less than the manufacturer's published value of 5~microns.

\articlefigure[scale=0.6]{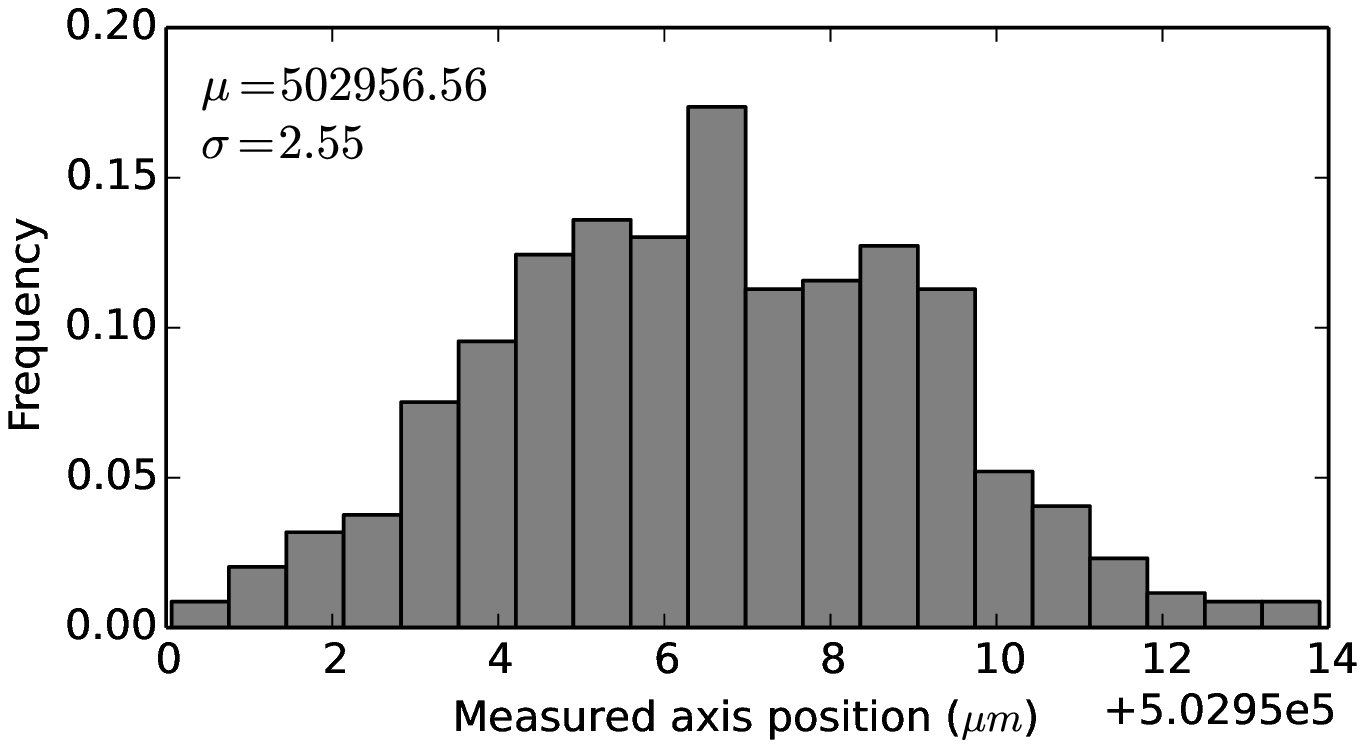}{plot_axis_rep}{Histogram plot of the actual position of the $y$-axis when instructed to go to the same position 500 times}

\altsubsubsection{Gripper release repeatability}
The gripper jaws were tested for repeatability when releasing fibers.  A fiber was picked up, placed back down, and released.  This was repeated 200 times, using two fiber buttons and two different locations along the $y$-axis.  The difference in fiber location between gripped and released states was measured for every cycle.

The results (Figure~\ref{plot_gripper_rep}) show a mean lateral shift of order 10~microns when releasing a button, mainly caused by the prototype field plate not being orthogonal to the $z$-axis of the robot gantry.  Crucially, however, the scatter of this offset is sub-micron.  This indicates that, as hoped, subtraction of the mean offset for each fiber button during positioning is feasible.

\articlefiguretwo{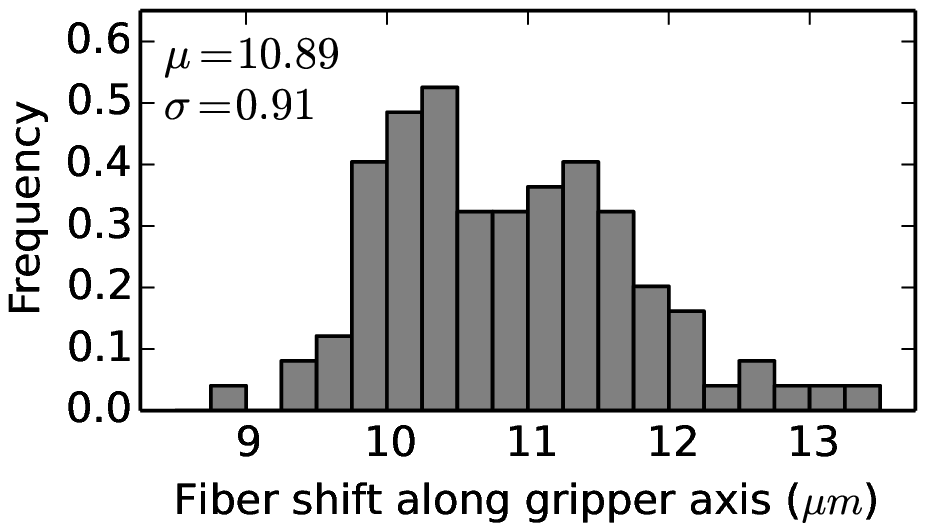}{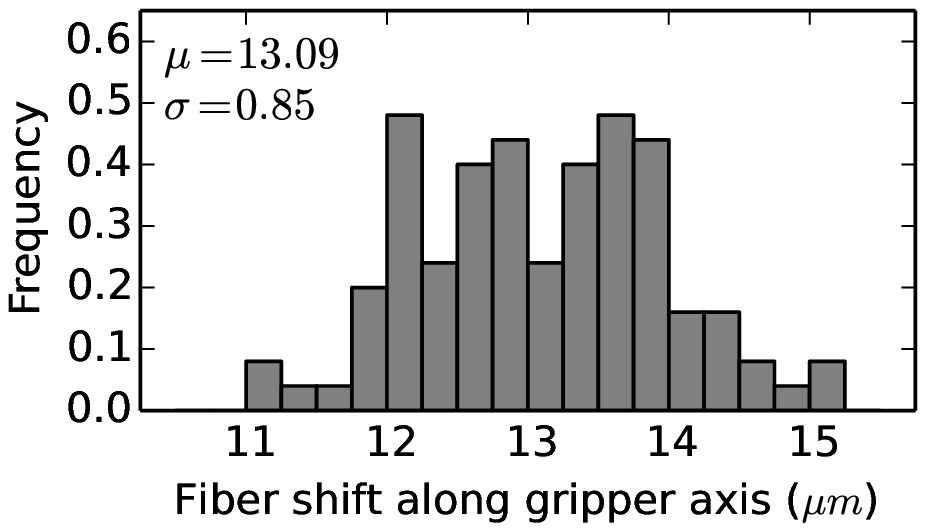}{plot_gripper_rep}{Histogram plots of the measured movement of a fiber when released by the robot gripper, for two different fibers at two different locations on the field plate}

\altsubsubsection{Timing estimates}
The expected field reconfiguration time for WEAVE has been significantly improved by parallelizing the pick-and-place operation.

Table~\ref{table_timing_estimates} shows an expected pick-and-place time of 4.5~seconds when each action in the procedure is done sequentially.  This would result in a total reconfiguration time of 79~minutes, exceeding the required maximum of 60~minutes (3.4~seconds per move).

By allowing some motor actions to begin before others have ended, the time spent waiting for moves to complete is significantly reduced.  Table~\ref{table_timing_estimates} lists the results of a test of this scheme, giving an expected time of 2.2~seconds per pick-and-place operation.  This reduces the estimated total reconfiguration time by more than half, to 39~minutes.

It is important to note that these estimates do not include any time spent waiting for the gantry structure to settle after a move, and that this cannot be tested until the full system has been assembled for the final instrument.  Since the total estimated positioning time now occupies only 65\% of that available, it seems reasonable to assume that mechanical settling delays will not result in a failure to meet the requirement.

\begin{table}[!ht]
\caption{A comparison of pick-and-place timings for the mean move distance of 205~mm. Safely overlapping moves significantly reduces the total positioning time.}
\smallskip
\begin{center}
{\small
\begin{tabular}{lll}
\tableline
\noalign{\smallskip}
    Operation requiring wait            & Mean wait, sequential (s)     & Mean wait, overlapped (s) \\
\noalign{\smallskip}
\tableline
\noalign{\smallskip}
    Movement of robot axes (pick)       & 1.9                           & 0.71                       \\
    Movement of gripper jaws (pick)     & 0.25                          & 0.17                       \\
    Fiber metrology (pick)              & 0.10                          & 0.10                       \\
    Movement of robot axes (place)      & 1.9                           & 0.86                       \\
    Movement of gripper jaws (place)    & 0.25                          & 0.25                       \\
    Fiber metrology (place)             & 0.10                          & 0.10                       \\
\noalign{\smallskip}
\tableline
\noalign{\smallskip}
    Entire pick-and-place operation     & 4.5                           & 2.2                        \\
\noalign{\smallskip}
\tableline\
\end{tabular}
\label{table_timing_estimates}
}
\end{center}
\end{table}

\section{Conclusions}
Testing of a prototype positioner has verified the published performance values for the COTS systems chosen for WEAVE.  The performance of the axis and gripper is significantly better than specified, indicating that the requirement for a maximum positioning error of 8.0~microns is achievable.

The estimated total reconfiguration time for WEAVE has been improved by parallelizing the pick-and-place operation.  Although mechanical settling times are not yet included, 45\% of the available time window is still available for such delays.  Therefore it is reasonable to believe that full fields can be reconfigured to the necessary accuracy within the allowed 60~minutes.

Positioner performance at different inclinations is accounted for by the use of counterweights on all gantry axes.  Significant changes in behavior are, therefore, not anticipated with telescope movement.  This will be confirmed in the next testing phase.

\end{document}